\documentclass{elsart}

\begin{document}

\begin{frontmatter}

\title{Effective interactions in medium heavy nuclei}

\author{T.\ Engeland}, 
\author{M.\ Hjorth-Jensen} and
\author{E.\ Osnes}
\address{Department of Physics,
         University of Oslo, N-0316 Oslo, Norway}

\maketitle

\begin{abstract}

We present a brief overview  
of microscopic nuclear structure approaches
to nuclei with 
$A\sim 100-132$. 
The emphasis is on the shell
model and theories for deriving effective 
interactions starting from the free interactions
between nucleons. New results for $^{105,106,107}$Sb are presented.  

\end{abstract}

\begin{keyword}
Shell model; Effective interactions
\end{keyword}

\end{frontmatter}

\section{Introduction}

Nuclei far from the line of $\beta$-stability are at present in
focus of the nuclear structure physics community. 
Considerable attention is being devoted to the
experimental and theoretical 
study of nuclei near $^{100}$Sn, from studies of the chain of Sn isotopes
up to 
$^{132}$Sn to e.g., nuclei near the proton drip line like $^{105,106}$Sb. 
Nuclei like
$^{105}$Sb and $^{109}$I have recently  been established as  ground-state 
proton emitters \cite{sb105,i109}.
The next to drip line nucleus for the antimony isotopes, $^{106}$Sb
with a proton separation energy of $\sim 400$ keV, was studied recently
in two experiments and a level scheme for the yrast states was proposed
in Ref.~\cite{sb106}. Similarly, detailed spectroscopy of $^{107}$Sb
has also been presented recently \cite{sb107}.

In this contribution we focus 
on selected aspects of  tin isotopes, with an emphasis
on the connection to the underlying nucleon-nucleon interaction
and large-scale shell-model studies with effective interactions.
The derivation of a shell-model effective interaction is briefly discussed
in section 2. 
Since the effective interactions employed in shell-model calculations 
are always the outcome of some truncations in the many-body expansion,
the shell model may then provide a useful testing ground for 
the various approximations made. Furthermore, the shell-model wave function
can be used to extract information on specific correlations in nuclei, such as
pairing correlations. This is discussed in section 3, while
section 4 contains new results from studies of Sb isotopes.
Concluding remarks are presented
in section 5.

\section{Effective interactions and the shell model}

Our scheme to obtain an effective two-body interaction for 
shell-model studies
starts with a free nucleon-nucleon  interaction $V$ which is
appropriate for nuclear physics at low and intermediate energies. 
In this work we will thus choose to work with the charge-dependent
version of the Bonn potential models, see Ref. \cite{cdbonn}.
The next step 
in our many-body scheme is to handle 
the fact that the repulsive core of the nucleon-nucleon potential $V$
is unsuitable for perturbative approaches. This problem is overcome
by introducing the reaction matrix $G$, which in 
a diagrammatic language represents  the sum over all
ladder type of diagrams. This sum is meant to renormalize
the repulsive short-range part of the interaction. The physical interpretation
is that the particles must interact with each other an infinite number
of times in order to produce a finite interaction. 
We calculate $G$ using the double-partioning scheme discussed
in e.g.,~Ref. \cite{hko95}.
Since the $G$-matrix represents just
the summation to all orders of particle-particle
ladder diagrams, there are obviously other terms which need to be included
in an effective interaction. Long-range effects represented by 
core-polarization terms are also needed.
In order to achieve this,  the $  G  $-matrix elements
are renormalized by the $\hat{Q}$-box method.
The $\hat{Q}$-box is made up of non-folded diagrams which are irreducible
and valence linked. Here we include all non-folded diagrams to third
order in $G$ \cite{hko95}.
Based on the $\hat{Q}$-box, we compute 
an effective interaction
$\tilde{H}$ in terms of the $\hat{Q}$-box,
 using the folded-diagram expansion method, 
see e.g., Ref.\  \cite{hko95} for further details.

The effective two-particle interaction can in turn be used in 
large-scale shell model
calculations.
The shell model problem requires the solution of a real symmetric
$n \times n$ matrix eigenvalue equation
$\tilde{H}\left | \Psi_k\right\rangle  = 
       E_k \left | \Psi_k\right\rangle$ ,
with $k = 1,\ldots, K$. 
At present our basic approach to 
finding solutions to this equation
is the Lanczos algorithm; an iterative method which gives the solution of
the lowest eigenstates. 
The technique is described in detail in Ref.\ \cite{whit77}. 

In our studies the shell-model space consists of the orbits 
$2s_{1/2}$, $1d_{5/2}$, $1d_{3/2}$, $0g_{7/2}$ and $0h_{11/2}$, for
both protons and neutrons. For the studies of the Sb isotopes we used
$^{100}$Sn as closed shell core, while for the heavy tin isotopes
we derived an effective interaction with $^{132}$Sn as closed shell core,
see e.g., Refs.~\cite{sb106,ehho98} for calculational details. 
The dimensionality
$n$ of the eigenvalue matrix $\tilde{H}$ is increasing
with increasing number of valence particles or holes. As an example,
for $^{116}$Sn 
the dimensionality of the hamiltonian matrix is of  
the order of  $n \sim \times 10^{8}$.

\section{Selected features of tin isotopes}

Of interest in this study is the fact that 
the chain of even tin isotopes from $^{102}$Sn to $^{130}$Sn 
exhibits a near constancy of the 
$2^+_1-0^+_1$ excitation energy, a constancy which can be related
to strong pairing correlations and the near degeneracy in energy 
of the relevant single particle orbits. As an example, we show the 
experimental\footnote{We will limit our discussion to even nuclei
from  $^{116}$Sn to $^{130}$Sn, since a qualitatively similar picture
is obtained from $^{102}$Sn to $^{116}$Sn.}
$2^+_1-0^+_1$ excitation energy 
from  $^{116}$Sn to $^{130}$Sn in Table \ref{tab:table1}. 
Our aim is to see whether partial waves which play a crucial
role in superfluidity of neutron star matter \cite{hh2000}, 
viz., $^1S_0$ and $^3P_2$, are equally
important in reproducing the near constant spacing in the chain
of even tin isotopes shown in  Table \ref{tab:table1}.

In order to test whether the $^1S_0$ and $^3P_2$ partial waves are equally
important in reproducing the near constant spacing in the chain
of even tin isotopes as they are for the superfluid properties of infinite 
neutron star matter,
we study four different approximations to the shell-model
effective interaction, viz.,
\begin{enumerate}
  \item Our best approach to the effective interaction, $V_{\mathrm{eff}}$, contains
        all one-body and two-body diagrams through third order in the $G$-matrix, 
        see Ref.\ \cite{ehho98}. 
  \item The effective interaction is given by the $G$-matrix only and inludes
        all partial waves up to $l=10$.
  \item We define an effective  interaction based on a $G$-matrix which now includes
        only the $^1S_0$ partial wave.
  \item Finally, we use an effective interaction based on a $G$-matrix which does
        not contain the  $^1S_0$ and $^3P_2$ partial waves, but all other waves
        up to $l=10$.  
\end{enumerate}
In all four cases the same NN interaction is used, viz., 
the CD-Bonn interaction described in Ref.\ \cite{cdbonn}.
Table \ref{tab:table1} lists the results.  
\begin{table}[hbt]
\begin{center}
\caption{ $2^+_1-0^+_1$ excitation energy for the 
even tin isotopes $^{130-116}$Sn for various approaches
to the effective interaction. See text for further details. 
Energies are given in MeV. }\footnotesize
\begin{tabular}{lcccccccc}\hline
 & {$^{116}$Sn} & {$^{118}$Sn} & {$^{120}$Sn} &{$^{122}$Sn} & {$^{124}$Sn} & {$^{126}$Sn} & {$^{128}$Sn} & {$^{130}$Sn} \\ \hline
Expt & 1.29 & 1.23 & 1.17 & 1.14 & 1.13 & 1.14 & 1.17 & 1.23 \\
$V_{\mathrm{eff}}$ & 1.17 & 1.15 & 1.14 & 1.15 & 1.14 & 1.21 & 1.28 & 1.46 \\
$G$-matrix &1.14 & 1.12& 1.07 & 0.99 & 0.99 & 0.98 & 0.98 & 0.97  \\
$^1S_0$ $G$-matrix &1.38 &1.36 &1.34 &1.30 & 1.25& 1.21 &1.19 &1.18 \\
No $^1S_0$ \& $^3P_2$ in $G$ &     &     &     &      &0.15 &-0.32 &0.02 &-0.21  \\\hline
\end{tabular}
\end{center}
\label{tab:table1}
\end{table}
We note from this table that the three first cases nearly produce a constant 
$2^+_1-0^+_1$ excitation energy, with our most optimal effective interaction
$V_{\mathrm{eff}}$ being closest the experimental data. The bare $G$-matrix
interaction, with no folded diagrams as well, results in a slightly more compressed
spacing. This is mainly due to the omission of the core-polarization 
diagrams which typically render the $J=0$ matrix elements more attractive.
Such diagrams are included in $V_{\mathrm{eff}}$. 
Including only the $^1S_0$ partial wave in the construction of the  $G$-matrix
(case 3),
yields in turn a somewhat larger spacing. This can again be understood from the
fact that a $G$-matrix constructed with this partial wave  
only does not receive contributions from any entirely repulsive partial wave.
It should be noted that our optimal interaction, as demonstrated in 
Ref.\ \cite{ehho98}, shows a rather good reproduction of the 
experimental spectra for both even and odd nuclei. Although the approximations
made in cases 2 and 3 produce an almost constant $2^+_1-0^+_1$ excitation energy,
they reproduce poorly the properties of odd nuclei and other 
excited states in the even Sn isotopes. 

However, the fact that the first three  approximations result in a such a good
reproduction of the  $2^+_1-0^+_1$ spacing may hint to the fact that the 
$^1S_0$ partial wave is of paramount importance. 
If we now turn the attention to case 4, i.e., we omit the
$^1S_0$ and $^3P_2$ partial waves in the construction of the $G$-matrix,
the results presented  in Table \ref{tab:table1} exhibit  a spectroscopic 
catastrophe. We do also not list eigenstates
with other quantum numbers. For e.g., $^{126}$Sn
the ground state is no longer a $0^+$ state, rather it carries $J=4^+$ while for $^{124}$Sn the ground state 
has $6^+$. The first $0^+$ state for this nucleus is given at an excitation
energy of $0.1$ MeV with respect to the $6^+$ ground state.
The general picture for other eigenstates is that of 
an extremely poor agreement
with data.  
Since the agreement is so poor, even the qualitative reproduction of the 
$2^+_1-0^+_1$ spacing, we defer from performing time-consuming shell-model
calculations for $^{116,118,120,122}$Sn.

\section{Shell model studies of the proton drip line nuclei 
$^{105,106,107}$Sb}

We present recent results for $^{105}$Sb, $^{106}$Sb and $^{107}$Sb in
Table 2. The calculations use $^{100}$Sn as
closed shell core with an effective 
interaction for the four, five or six valence neutrons and one valence proton
based on the CD-Bonn nucleon-nucleon interaction \cite{cdbonn}.
The experimental spin assignements for $^{105}$Sb and
$^{106}$Sb are tentative. 
There are also many more theoretical states than 
reported in the enclosed table. 

The high spin level scheme of $^{105}$Sb resembles the level scheme of
$^{107}$Sb up to $J=19/2$ \cite{sb107}. 
When compared to $^{107}$Sb the $^{105}$Sb
level scheme shows similar trends as when going from $^{106}$Sn to
$^{104}$Sn. That means that coupling a $d_{5/2}$ proton to a 
$^{104}$Sn core is appropriate to describe the observed states.
The calculation favors $J^{\pi}$=5/2$^+$ for the ground state in
agreement with the suggestion from proton decay data.
In this state the valence proton is mainly in the $d_{5/2}$
orbit and the two neutron pairs are almost evenly distributed
over the $d_{5/2}$ and $g_{7/2}$ neutron orbits. The situation
is very similar in the 9/2$^+$ and 13/2$^+$ states, while 
the $\nu$$g_{5/2}^3$$g_{7/2}^1$ configuration exhausts the
largest parts of the wave functions of the 15/2$^+$ and 17/2$^+$ 
states. The neutron part of the wave function of the 19/2$^+$
state is almost identical to the 17/2$^+$ state. However,
since 17/2$^+$ is the maximum spin for the 
$\pi$$g_{5/2}^1$$\nu$$g_{5/2}^1$$g_{7/2}^1$ configuration,
the odd proton resides almost exclusively in the $g_{7/2}$
orbit in the 19/2$^+$ state. 
For proton degrees of freedom
the $s_{1/2}$, $d_{3/2}$ and $h_{11/2}$ single-particle 
orbits give essentially negligible
contributions to the wave functions and the energies of the excited
states, as expected. For neutrons, although
the single-particle distribution for a given state is also negligible,
these orbits are important for a good describtion of the energy spectrum,
as also demonstrated in large-scale shell-model calculations of
tin isotopes \cite{ehho98}.
Similar picturer applies to $^{106}$Sb and $^{107}$Sb as well, 
see Refs.~\cite{sb106,sb107}.
The wave functions for the various states are to a large extent
dominated by the $g_{7/2}$ and $d_{5/2}$ single-particle orbits
for neutrons ($\nu$)
and the $d_{5/2}$ single-particle
orbit for protons ($\pi$). 
The $\nu g_{7/2}$ and $\nu d_{5/2}$ single-particle orbits represent
in general more than $\sim 90\%$ of the total neutron single-particle
occupancy, while the  $\pi d_{5/2}$ single-particle orbits stands for 
$\sim 80-90\%$ of the proton single-particle occupancy. The other single-particle
orbits play  an almost negligible role in the structure of the wave functions.
\begin{table}[hbt]
\begin{center}
\caption{ Low-lying states of $^{105,106,107}$Sb, theory and experiment.
Energies in MeV. }\footnotesize
\begin{tabular}{ccc|ccc|ccc}
\hline
\multicolumn{3}{c|}{ $^{105}$Sb} & \multicolumn{3}{c|}{ $^{106}$Sb}& \multicolumn{3}{c}{ $^{107}$Sb} \\ 
{$J^{\pi}_i$} & {Exp} & {Theory} & 
{$J^{\pi}_i$} & {Exp} & {Theory} & 
{$J^{\pi}_i$} & {Exp} & {Theory} \\
\hline 
$5/2^{+}$ & 0 & 0 & $2^{+}$ & 0 & 0 & $5/2^{+}$ & 0 & 0 \\
$9/2^{+}$ & 1.22 & 1.22 & $4^{+}$ & 0.10 & 0.25 & $7/2^{+}$ & 0.77 & 0.69 \\
$13/2^{+}$ & 1.84 & 1.94 & $5^{+}$ & 0.32 & 0.54 & $9/2^{+}$ & 1.06 & 1.08  \\
$15/2^{+}$ & 2.21 & 2.10 & $6^{+}$ & 0.44 & 0.66 & $11/2^{+}$ & 1.79 & 1.80 \\
$17/2^{+}$ & 2.50 & 2.41 & $7^{+}$ & 0.89 & 1.34 & $13/2^{+}$ & 1.90 & 1.94  \\
$19/2^{+}$ & 2.99 & 2.94 & $8^{+}$ & 1.53 & 1.78 &  $15/2^{+}$ & 2.24 & 2.38\\
$23/2^{+}$ & 3.73 & 4.09 & $10^{+}$ & 2.26 & 2.57 & $17/2^{+}$ & 2.75 & 2.83  \\\hline
\end{tabular}
\end{center}
\end{table}

\section{Conclusions}

In summary, shell-model calculations with realistic effective
interactions of the newly reported low-lying
yrast states of  the proton drip line nuclei $^{105,106,107}$Sb, reproduce well
the experimental data. Since the wave functions of the various states are to a large
extent dominated by neutronic degrees of freedom and neutrons are well
bound with a separation energy of $\sim 8$ MeV, this may explain why a shell-model
calculation, within a restricted model space for a system close to the proton drip line,
gives a satisfactory agreement with the data.  

For the Sn isotopes we have shown that
the $^1S_0$ and $^3P_2$ partial waves, which are crucial for our
understanding of superfluidity in neutron star matter, are equally
important in order to reproduce the 
$2^+_1-0^+_1$ excitation energy of
the even Sn isotopes. Omitting these waves, especially the  $^1S_0$ wave,
results in a spectrum which has essentially no correspondence with experiment.

Further analysis of nuclei such as Ag, Cd, In near $A\sim 100$ and 
Sn isotopes near $A\sim 132$ are in progress.

We are much indebted to Cyrus Baktash, David Dean, Hubert Grawe  
and Matej Lipoglav\v{s}ek 
for many discussions on properties of nuclei near $A\sim 100$.

\end{document}